\documentclass{article}
\usepackage{spconf,amsmath,epsfig,comment,url}

\usepackage{epsfig,comment,cite,latexsym,xcolor,balance}
\def\D{\mbox{\rm D}}

\newcommand{\bm}[1]{\mbox{\boldmath$#1$\unboldmath}}

\newcommand{\beqn}{\begin{eqnarray}}
\newcommand{\eeqn}{\end{eqnarray}}
\newcommand{\beq}{\begin{eqnarray*}}
\newcommand{\eeq}{\end{eqnarray*}}
\newcommand{\tr}{\mbox{tr}}  % trace
\newcommand{\z}{\mbox{{\boldmath${z}$\unboldmath}}}  % bold z
\newcommand{\0}{\mbox{{\boldmath${0}$\unboldmath}}}  % bold 0
\newcommand{\I}{\mbox{{\boldmath${I}$\unboldmath}}}  % bold I
\newcommand{\A}{\mbox{{\boldmath${A}$\unboldmath}}}  % bold C
\newcommand{\C}{\mbox{{\boldmath${C}$\unboldmath}}}  % bold C
\newcommand{\X}{\mbox{{\boldmath${X}$\unboldmath}}}  % bold X
\newcommand{\Y}{\mbox{{\boldmath${Y}$\unboldmath}}}  % bold Y
\newcommand{\Z}{\mbox{{\boldmath${Z}$\unboldmath}}}  % bold Z
\newcommand{\W}{\mbox{{\boldmath${W}$\unboldmath}}}  % bold W
\newcommand{\w}{\mbox{{\boldmath${w}$\unboldmath}}}  % bold w
\newcommand{\aC}{\langle\C\rangle}  % bold <C>

\title{\LARGE \bfseries \sf Generation of Sample Complex Wishart Distributed Matrices and\\ \vspace{3pt} Change Detection in Polarimetric SAR Data}

\name{\sf Allan A. Nielsen$^a$, Henning Skriver$^b$ and Knut Conradsen$^a$\thanks{Allan Nielsen: phone +45 4525 3425, e-mail alan@dtu.dk, homepage https://people.compute.dtu.dk/alan. Presented at the International Conference on Digital Image and Signal Processing (DISP), Oxford, UK, April 2019.}}
\address{ \vspace{2pt}
\sf Technical University of Denmark\\ \vspace{2pt}
\sf $^a$DTU Compute -- Applied Mathematics and Computer Science\\ \vspace{2pt}
\sf $^b$DTU Space -- National Space Institute\\ \vspace{2pt}
\sf DK-2800 Kgs.\ Lyngby, Denmark}

\begin{document}
\ninept

\maketitle

\thispagestyle{empty}

\urlstyle{same}

%\tableofcontents

%\begin{abstract}
{\noindent \bfseries ABSTRACT}

\noindent The complex Wishart distribution is used to describe synthetic aperture radar (SAR) data
in the so-called covariance matrix representation.
We give a test statistic for detection of differences between two instances in this distribution and an associated probability measure.
Generated complex Wishart distributed covariance matrices
are used to show that this test statistic and the probability measure in situations with no differences follow the expected distributions.\\
%\end{abstract}

%\begin{keywords}
{\noindent \bfseries Keywords \vspace{\baselineskip}}
\vspace{-\baselineskip}

\noindent Polarimetric SAR data,
dual polarization SAR data,
test statistic,\\
change
detection,
target detection.
%\noindent Remote sensing, data generation, test statistic for equality
%Polarimetric SAR.
%\end{keywords}

\section{\hspace{1em} Introduction}
\vspace{-0.9\baselineskip}
In this paper we study the behaviour of the test statistic and the associated probability measure for detection of differences such as change over time in polarimetric synthetic aperture radar (SAR) data in the covariance (or equivalently the coherency) matrix formulation \cite{zylulaby90} described in \cite{wishartchange2003,wishartedge2003}.
We study the test stastistic and the probability as functions of the equivalent number of looks (ENL).
To emulate polarimetric data (from for example Radarsat-2, ALOS or TerraSAR-X) and dual polarimetry data (from for example Sentinel-1 or COSMO-SkyMed), we generate many outcomes of both 3$\times$3 and 2$\times$2 Hermitian, positive definite matrices, as well as 2$\times$2 diagonal-only matrices with the same parameters (only, we let the ENL vary).
In a change detection setting, this emulates the situation of no change between two time points.
We therefore know the distributions of the test statistic and the probability, see \cite{wishartchange2003,wishartJSTARS2015} for details.
%We simulate 
%This is to emulate polarimetric data (from for example Radarsat-2) and dual polarimetry data (from for example Sentinel-1).

\begin{comment}
This paper describes the generation of samples of complex Wishart distributed Hermitian, positive definite matrices
\cite{odell1966}.
We apply this to simulated no-change in a change detection
setting in bi-temporal
(polarimetric) synthetic aperture radar (SAR) data
by means of methods first described in
\cite{wishartchange2003}.
%\cite{wishartchange2003,canty3rd2014,wishartJSTARS2015,wishartSeveralRj2016,wishartCJRS2017}.
We simulate both 3$\times$3 Hermitian, positive definite matrices, 2$\times$2 Hermitian, positive definite matrices, and
2$\times$2 diagonal-only matrices.
This is to emulate polarimetric data (from for example Radarsat-2) and dual polarimetry data (from for example Sentinel-1).
\end{comment}

\section{\hspace{1em} Synthetic Aperture Radar}
\vspace{-0.9\baselineskip}
A fully polarimetric SAR measures the 2$\times$2 complex so-called
scattering matrix at each resolution cell on the ground
\beq
\bm{S} & = &
\left[ \begin{array}{cc}
S_{hh} & S_{hv}\\
S_{vh} & S_{vv}\\
\end{array} \right]
.
\eeq
The scattering matrix relates the incident and the scattered
electric fields \cite{zylulaby90}. If $S_{rt}$ denotes the complex scattering amplitude
for receive and transmit polarizations ($r, t \in \{h,v\}$ for
horizontal and vertical polarizations), then reciprocity, which
normally applies to natural targets, gives $S_{hv} = S_{vh}$ (in the
backscattering direction using the backscattering alignment
convention) \cite{zylulaby90}. Assuming reciprocity, the scattering matrix
is represented by the three-component complex target vector
$\bm{s} = [S_{hh} \ S_{hv} \ S_{vv}]^T$ , where $^T$ means transpose.
The inherent speckle in the SAR data can be reduced by
spatial averaging at the expense of spatial resolution. In this so-called
multilook case a more appropriate representation of the
backscattered signal is the covariance matrix in which the average
properties of a group of resolution cells can be expressed
in a single matrix formed by the outer products of the averaged
target vectors. The average covariance matrix is defined as
\cite{zylulaby90}
\beq
\label{eq:covmaonct}
%\langle \C \rangle_{\mathit full} & = &
\langle \C \rangle & = &
\left[ \begin{array}{ccc}
\langle S_{hh}S_{hh}^* \rangle & \langle S_{hh}S_{hv}^* \rangle & \langle S_{hh}S_{vv}^* \rangle \\
\langle S_{hv}S_{hh}^* \rangle & \langle S_{hv}S_{hv}^* \rangle & \langle S_{hv}S_{vv}^* \rangle \\
\langle S_{vv}S_{hh}^* \rangle & \langle S_{vv}S_{hv}^* \rangle & \langle S_{vv}S_{vv}^* \rangle \\
\end{array} \right]
\eeq
where $\langle \cdot \rangle$ denotes
ensemble
averaging and $^*$ denotes complex conjugation.

\vspace{0.9\baselineskip}\noindent
Spaceborne instruments often transmit only one polarization,
say vertical, and receive both polarizations giving rise
to dual polarization data.
In the covariance matrix representation each pixel at each time point is a matrix
\beq
\label{eq:covmatdual}
\langle \C \rangle_{\mathit dual} & = &
\left[ \begin{array}{cc}
\langle S_{vv}S_{vv}^* \rangle & \langle S_{vv}S_{vh}^* \rangle \\
\langle S_{vh}S_{vv}^* \rangle & \langle S_{vh}S_{vh}^* \rangle \\
\end{array} \right].
\eeq
Sometimes we have the diagonal elements only.

\subsection{\hspace{1em} Test for Equality of Two Complex Covariance Matrices}
\label{sec:twocovmats}
\vspace{-0.6\baselineskip}
To test whether two complex variance-covariance matrices
$\bm{\Sigma}_x$ and
$\bm{\Sigma}_y$ (both $p$ by $p$)
are equal, i.e., to test the so-called null hypothesis
% $H_0$
\beq
H_0: \ \bm{\Sigma}_x = \bm{\Sigma}_y
\eeq
against the alternative,
$H_1: \ \bm{\Sigma}_x \neq \bm{\Sigma}_y$,
the following
test statistic applies
(for the real case see \cite{anderson2003})
\beq
\label{eq:Q}
%Q &=& \frac{(m+n)^{p(m+n)}}{m^{pm}n^{pn}} \frac{|\X|^m|\Y|^n}{|\X+\Y|^{m+n}}.
Q &=& \frac{(m+n)^{p(m+n)}}{m^{pm}n^{pn}} \frac{|m \aC_x|^m|n \aC_y|^n}{|m \aC_x+n \aC_y|^{m+n}}.
\eeq
Here
%the $\X = m \hat{\bm{\Sigma}}_x = m \aC_x$
%and $\Y = n \hat{\bm{\Sigma}}_y = n \aC_y$
$m \aC_x = m \hat{\bm{\Sigma}}_x $
and $n \aC_y = n \hat{\bm{\Sigma}}_y$
follow the complex Wishart distribution,
%i.e.,  $\X \sim W_C(p, m, \bm{\Sigma}_x)$, and
%$\Y \sim W_C(p, n, \bm{\Sigma}_y)$.
i.e.,  $m \aC_x \sim W_C(p, m, \bm{\Sigma}_x)$, and
$n \aC_y \sim W_C(p, n, \bm{\Sigma}_y)$.
$Q \in [0,1]$ with $Q=1$ for equality.

\vspace{0.9\baselineskip}\noindent
For the logarithm of the test statistic we get
\beqn
\ln Q &=& \nonumber
p \left\{ (m+n) \ln (m+n) - m \ln m - n \ln n \right\}\\ \nonumber
&& {}+ m \ln |m \aC_x| + n \ln |n \aC_y|\\ \nonumber
&& {}- (m+n) \ln |m \aC_x+n \aC_y|.
%&& {}+ m \ln |\X| + n \ln |\Y| -(m+n) \ln |\X+\Y|.
\eeqn
If
\beq
\rho &=& 1 - \frac{2p^2-1}{6p}\left( \frac{1}{m}+\frac{1}{n}-\frac{1}{m+n} \right)
\eeq
and
\beqn
\omega_2 &=& \nonumber
-\frac{p^2}{4} \left( 1-\frac{1}{\rho} \right)^2 + \frac{p^2(p^2-1)}{24\rho^2} \\ \nonumber
&& {} \cdot \left( \frac{1}{m^2}+\frac{1}{n^2}-\frac{1}{(m+n)^2} \right)
,
\eeqn
and if the observed value of $-2 \rho \ln Q$ is $z=-2 \rho \ln q_{obs}$, then the probability of finding a smaller value of
$-2 \rho \ln Q$ is
\beq
\label{eq:P}
\lefteqn{P\{-2 \rho \ln Q \leq z\} \ \simeq \ P\{\chi^2(f) \leq z\}}
	\\
&& {}+ \omega_2 [ P\{\chi^2(f+4) \leq z\} - P\{\chi^2(f) \leq z\} ]
\nonumber
.
\eeq
$1 - P\{-2 \rho \ln Q \leq z\} = P\{Q < q_{\rm obs} \}$ is the probability of equality, i.e.,
the probability of no change.
$f$ is the number of degrees of freedom.
For the polarimetric case, $f=3^2=9$, for dual polarization $f=2^2=4$, and for dual polarization with diagonal elements only,
$f=1^2+1^2=2$. 

\vspace{0.9\baselineskip}\noindent
These expressions are an improvement of the traditional test statistic $-2 \ln Q$ and the associated $P$
\beq
\label{eq:P0}
{P\{-2 \ln Q \leq -2 \ln q_{obs} \} \ \simeq \ P\{\chi^2(f) \leq z\}}
.
\eeq

\section{\hspace{1em} Generation of Complex Wishart Distributed Matrices}
\vspace{-0.9\baselineskip}
We wish to look into the distributions of $-2 \rho \ln Q$ and the associated $P$ for no difference or no change situations.
Therefore,
in this section we describe the generation of samples of complex Wishart distributed Hermitian, positive definite matrices
\cite{odell1966} where we can control the parameters $p$, $m$ (and $n$), and \bm{\Sigma}.
% in just that no difference situation.

\subsection{\hspace{1em} The complex multivariate normal distribution}
\vspace{-0.6\baselineskip}
A $p$-dimensional random vector $\Z \in C^p$
follows a complex multivariate normal distribution with
mean \0\ and dispersion matrix (vari\-ance-covariance matrix) \bm{\Sigma}, i.e.,
$\Z \sim N_C(\0,\bm{\Sigma})$
if the frequency function is
\beq
f(\z) = \frac{1}{\pi^p |\bm{\Sigma}|} \exp(-\z^H \bm{\Sigma}^{-1} \z)
\eeq
where the superscript $^H$ denotes the complex conjugate and transpose.

\vspace{0.9\baselineskip}\noindent
The dispersion matrix \bm{\Sigma} is Hermitian and positive definite.
We may write it on the form
\beq
\bm{\Sigma} =
%\left[ \begin{array}{cccc}
%\sigma_1^2 & (\alpha_{12} + i \beta_{12}) \sigma_1 \sigma_1 & (\alpha_{13} + i \beta_{13}) \sigma_1 \sigma_3 & \cdots\\
%(\alpha_{12} - i \beta_{12}) \sigma_1 \sigma_1 & \sigma_2^2 & (\alpha_{23} + i \beta_{23}) \sigma_2 \sigma_3 & \cdots\\
%(\alpha_{13} - i \beta_{13}) \sigma_1 \sigma_3 & (\alpha_{23} - i \beta_{23}) \sigma_2 \sigma_3 & \sigma_3^2 & \cdots\\
%\cdots & \cdots & \cdots & \ddots\\
%\end{array} \right]
\left[ \begin{array}{ccc}
\sigma_1^2 & (\alpha_{12} + i \beta_{12}) \sigma_1 \sigma_2 & \cdots\\
(\alpha_{12} - i \beta_{12}) \sigma_1 \sigma_2 & \sigma_2^2 &  \cdots\\
(\alpha_{13} - i \beta_{13}) \sigma_1 \sigma_3 &  (\alpha_{23} - i \beta_{23}) \sigma_2 \sigma_3 & \cdots\\
\vdots & \vdots  & \ddots\\
\end{array} \right].
\eeq
We can write this as $\bm{\Sigma} = \mathit{Re}(\bm{\Sigma}) + i \mathit{Im}(\bm{\Sigma})$.

\vspace{0.9\baselineskip}\noindent
If $\Z = \X + i \Y$ then the 2$p$-dimensional, real valued vector $[\X \ \ \Y]^T$
(where the superscript $^T$ denotes transpose)
has the variance-covariance matrix
\beq
%\D \left[ \X \ \Y \right] =
\D \left[ \begin{array}{r} \X \\ \Y\\ \end{array} \right] =
\frac{1}{2}\left[\begin{array}{rr}
\mathit{\mathit{Re}}(\bm{\Sigma}) & -\mathit{Im}(\bm{\Sigma})\\
\mathit{Im}(\bm{\Sigma}) &  \mathit{\mathit{Re}}(\bm{\Sigma})\\
\end{array} \right].
\eeq

\subsection{\hspace{1em} The complex Wishart distribution}
\vspace{-0.6\baselineskip}
A Hermitian, positive definite random matrix \W\ follows a complex Wishart distribution, i.e.,
$\W \sim W_C(p,n,\bm{\Sigma})$
if the frequency function is
\beq
f(\w) = \frac{1}{\Gamma_p(n)} \frac{1}{|\bm{\Sigma}|^n} |\w|^n \exp(\tr (\bm{\Sigma}^{-1} \w))
\eeq
where $\tr$ denotes the trace and
\beq
\Gamma_p(n) = \pi^{p(p-1)/2} \prod_{j=1}^p \Gamma(n-j+1)
\eeq
where $\Gamma$ is the usual gamma function.

\subsection{\hspace{1em} The empirical variance-covariance matrix for normal samples}
\vspace{-0.6\baselineskip}
Let $\Z_1, \ldots, \Z_n$ be independent $N_C(\0,\bm{\Sigma})$-distributed random variables.
Let $\hat{\bm{\Sigma}}$
be the maximum
likelihood estimate of \bm{\Sigma}.
Then
\beq
n \hat{\bm{\Sigma}} = \sum_{j=1}^n \Z_j \Z_j^H \sim W_C(p,n,\bm{\Sigma}).
\eeq

\subsection{\hspace{1em} Linear transformations}
\vspace{-0.6\baselineskip}
If $\Z \sim N_C(\0,\bm{\Sigma})$ then the linearly transformed variable $\A \Z$
where \A\ is a matrix of constants
will again be normal, and
$\A \Z \sim N_C(\0, \\
\A \bm{\Sigma} \A^H)$.
If $\W \sim W_C(p,n,\bm{\Sigma})$ then $\A \W \A^H \sim W_C(p,n,\A \bm{\Sigma} \A^H)$.

\subsection{\hspace{1em} Generation of \bm{W_C(p,n,\bm{\I})} and \bm{W_C(p,n,\bm{\Sigma})} distributed random variables}
\vspace{-0.6\baselineskip}
We consider observations $\Z_j = \X_j + i \Y_j, j=1,\ldots,n$ and organize the real and the imaginary parts
in the data matrices
\beq
\X =
\left[\begin{array}{c}
\X_1^T\\
\vdots\\
\X_n^T\\
\end{array} \right]
=
\left[\begin{array}{ccc}
X_{11} & \cdots & X_{1p}\\
\vdots &  & \vdots\\
X_{n1} & \cdots & X_{np}\\
\end{array} \right]
\eeq
\beq
\Y =
\left[\begin{array}{c}
\Y_1^T\\
\vdots\\
\Y_n^T\\
\end{array} \right]
=
\left[\begin{array}{ccc}
Y_{11} & \cdots & Y_{1p}\\
\vdots &  & \vdots\\
Y_{n1} & \cdots & Y_{np}\\
\end{array} \right]
\eeq
and we assume that all $X$ and $Y$ variables are independent and $N(0,1)$-distributed.
Then
\beq
\bm{S} = \frac{1}{2} \{\X^T\X + \Y^T\Y - i (\X^T\Y - \Y^T\X)\} \sim W_C(p,n,\bm{\I}).
\eeq

\vspace{0.9\baselineskip}\noindent
If \C\ is the Cholesky factorization of the
%arbitrary
Hermitian, positive definite matrix \bm{\Sigma}, i.e.,
$\bm{\Sigma} = \C \C^H$,
then
\beq
\W = \C \bm{S} \C^H \sim W_C(p,n,\bm{\Sigma}).
\eeq

\newpage

\section{\hspace{1em} Experiments}
\vspace{-0.9\baselineskip}
To emulate
polarimetric and
dual polarization data from
%for example Radarsat-2 and different airborne SAR systems, and dual polarization data from Sentinel-1,
real SAR systems such as Radarsat-2 (http://www.asc-csa.gc.ca/eng/satellites/ radarsat2) and Sentinel-1(https://sentinel.esa.int/web/sentinel/missions/
sentinel-1),
%real SAR systems such as Radarsat-2\footnote{\hspace{1pt}http://www.asc-csa.gc.ca/eng/satellites/radarsat2} and Sentinel-1\footnote{\hspace{1pt}https://sentinel.esa.int/web/sentinel/missions/sentinel-1},
we use the above to generate two instances of samples of complex Wishart distributed 3$\times$3, 2$\times$2 as well as 2$\times$2 diagonal-only matrices.
In this latter case, the matrices are not complex (and they are not Wishart distributed).
However, in
\cite{wishartchange2003}
we show that also in this case we can use the tools developed there for detection of differences such as change over time in (polarimetric) SAR data, see also
\cite{canty3rd2014,wishartJSTARS2015,wishartSeveralRj2016,wishartCJRS2017}.

\vspace{0.9\baselineskip}\noindent
Two experiments are performed, one with very different equivalent numbers of looks, ENL, 10 and 100.
The other experiment deals with combinations of ENL up to 20 in the two instances.

\subsection{\hspace{1em} Very different ENL}
\vspace{-0.6\baselineskip}
In a first experiment,
we generate 128k samples of independent complex Wishart distributed Hermitian, positive definite matrices
for each of two instances with the same \bm{\Sigma}, one with $L_1=100$ looks, one with $L_2=10$ looks.
%This corresponds to no change between two time points in a change detection setting
This corresponds to no change in a change detection setting
or to no target in a target detection setting \cite{targetdetsar2013}
where the number of looks for a centre pixel and its surrounding pixels are very different.
With no change or target, we know the asymptotic distribution of the test statistic $-2 \rho \ln Q$
and the associated $P$-value.

\begin{figure}
\centering
\includegraphics[width=0.95\columnwidth]{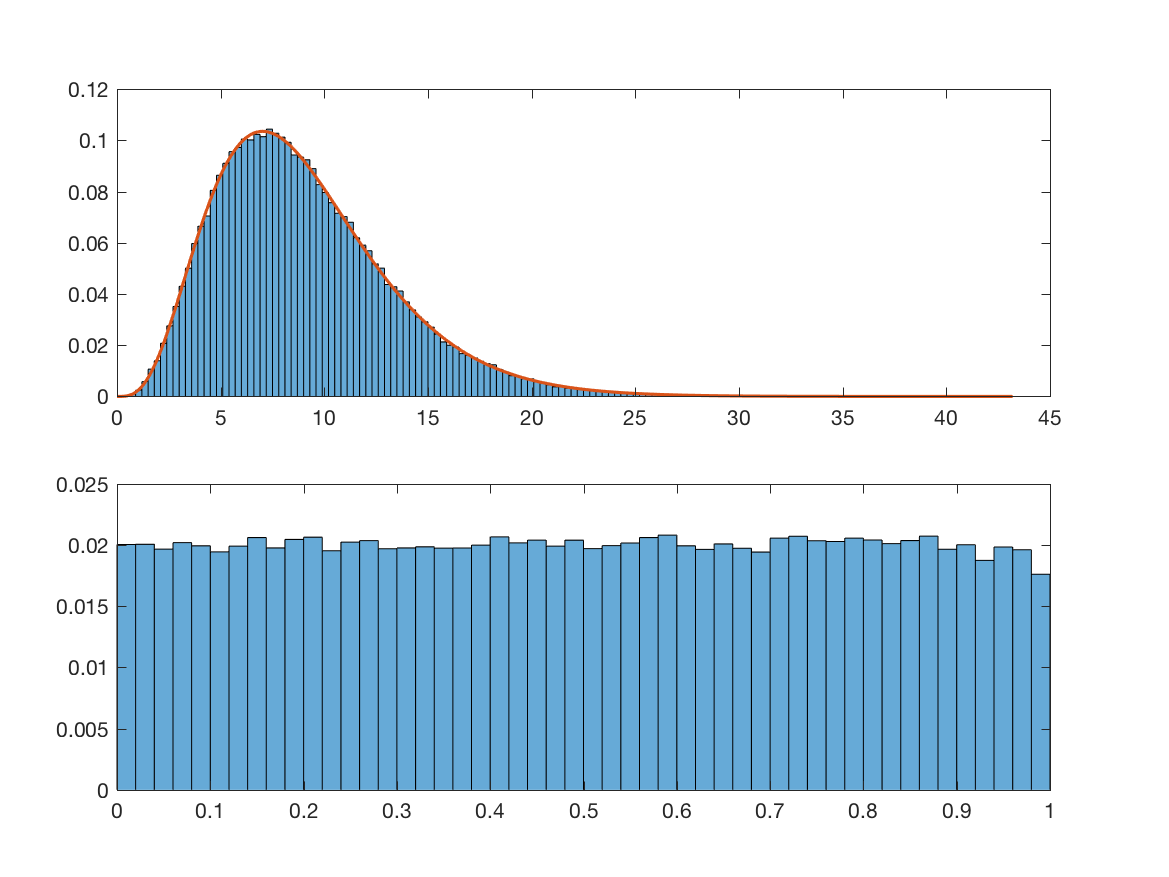}
\caption{\label{fig:histm2rlnQ3x3}Histograms for the no target situation with $L_1=100$ and $L_2=10$
for 3$\times$3 matrices,
$-2 \rho \ln Q$ and the theoretical distribution in red (top), and $P$ (bottom).}
%\end{figure}
\vspace{-0.2\baselineskip}
%\begin{figure}
\centering
\includegraphics[width=0.95\columnwidth]{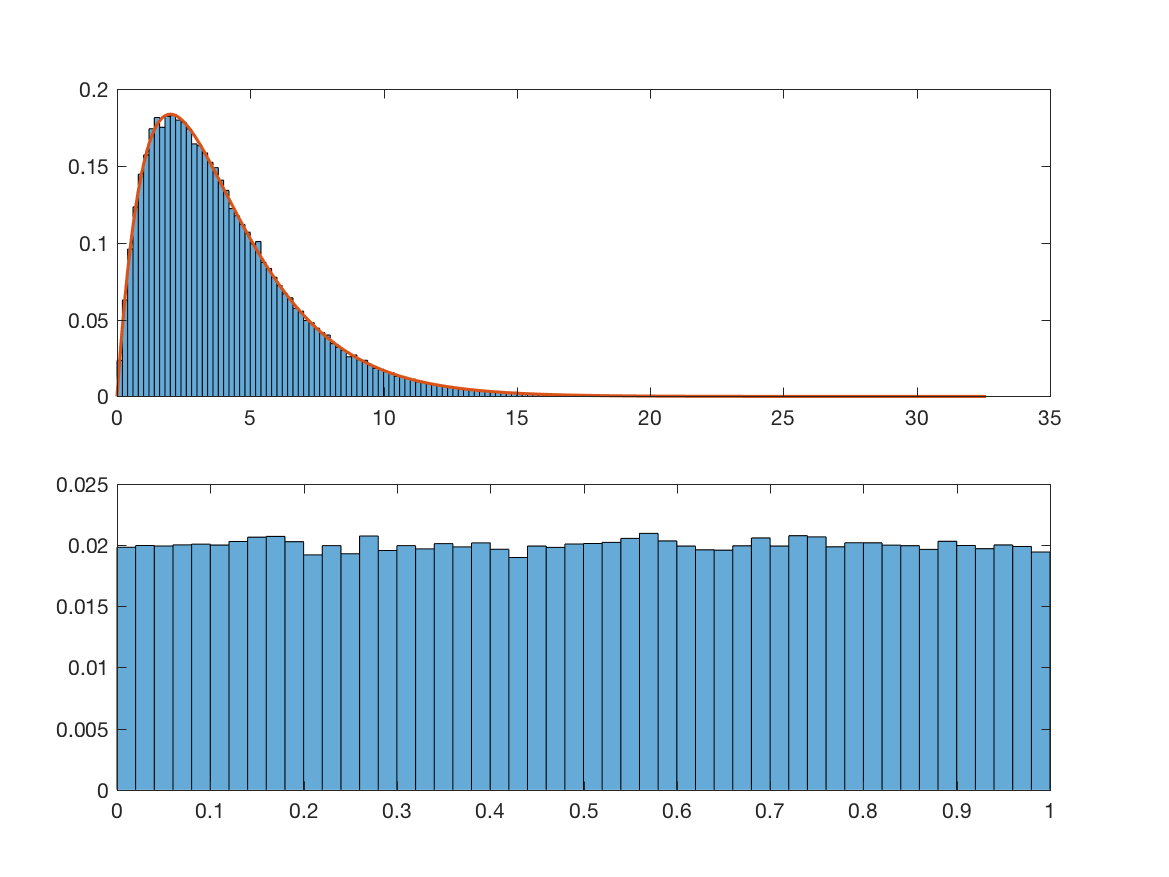}
\caption{\label{fig:histm2rlnQ2x2}Histograms for the no target situation with $L_1=100$ and $L_2=10$
for 2$\times$2 matrices,
$-2 \rho \ln Q$ and the theoretical distribution in red (top), and $P$ (bottom).}
%\end{figure}
\vspace{-0.2\baselineskip}
%\begin{figure}
\centering
\includegraphics[width=0.95\columnwidth]{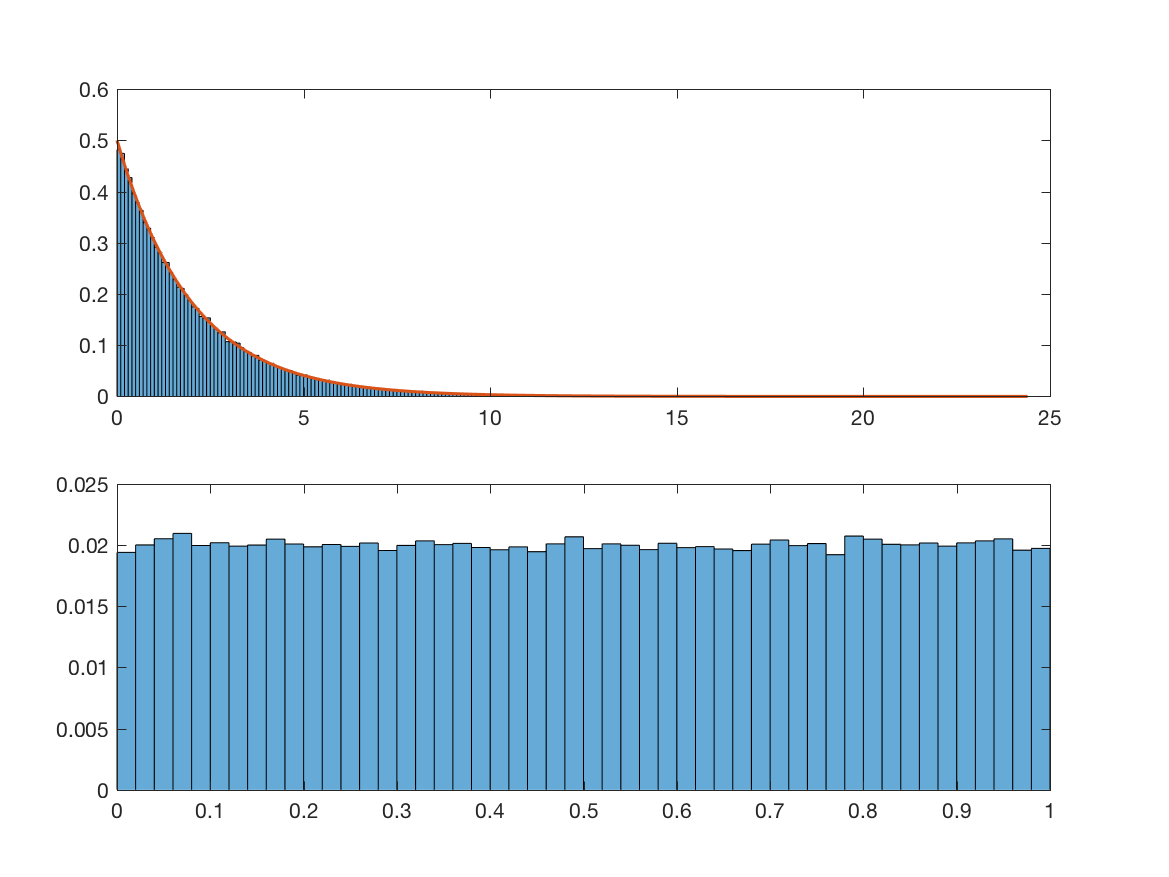}
\caption{\label{fig:histm2rlnQ2x2diag}Histograms for the no target situation with $L_1=100$ and $L_2=10$
for 2$\times$2 diagonal-only matrices,
$-2 \rho \ln Q$ and the theoretical distribution in red (top), and $P$ (bottom).}
\end{figure}

\vspace{0.9\baselineskip}\noindent
Figure~\ref{fig:histm2rlnQ3x3} shows histograms for this no target situation
%(where the number of looks are very different)
for 3$\times$3 matrices,
$-2 \rho \ln Q$ and the theoretical distribution in red (top), and $P$ (bottom).
%The histograms very beautifully show the expected behaviour for no-change including the uniform distribution of $P$.

\vspace{0.9\baselineskip}\noindent
Figure~\ref{fig:histm2rlnQ2x2} shows histograms for the no target situation
%(where the number of looks are very different)
for 2$\times$2 matrices,
$-2 \rho \ln Q$ and the theoretical distribution in red (top), and $P$ (bottom).
%The histograms very beautifully show the expected behaviour for no-change including the uniform distribution of $P$.

\vspace{0.9\baselineskip}\noindent
Figure~\ref{fig:histm2rlnQ2x2diag} shows histograms for the no target situation
%(where the number of looks are very different)
for 2$\times$2 diagonal-only matrices,
$-2 \rho \ln Q$ and the theoretical distribution in red (top), and $P$ (bottom).
%The histograms very beautifully show the expected behaviour for no-change including the uniform distribution of $P$.

\vspace{0.9\baselineskip}\noindent
In all three cases depicted in Figures~\ref{fig:histm2rlnQ3x3}, \ref{fig:histm2rlnQ2x2} and \ref{fig:histm2rlnQ2x2diag}
the histograms very beautifully show the expected behaviour for no target:
the combination of $\chi^2$ distributions of $-2 \rho \ln Q$ and the uniform distribution of $P$.

\subsection{\hspace{1em} ENL up to 20}
\vspace{-0.6\baselineskip}
In a second experiment,
we generate
%10,000
128k
samples of independent complex Wishart distributed data for each of two instances with the same \bm{\Sigma}.
Results are described in detail in the subsections below,
Again, this experiment emulates the situation with no change between two time points in a change detection setting.
Ideally, of course, there is no change and we therefore know the asymptotic distributions of the test statistic $-2 \rho \ln Q$
(and the simpler $-2 \ln Q$) and the associated $P$-value.
%We therefore also know the mean values
The calculations are carried out for a number of combinations of ENL for both ``time points''.
In the subsections below, we show plots of the mean value of the
%10,000
128k
samples of $-2 \rho \ln Q$ (and $-2 \ln Q$) and the associated $P$-values as functions of ENL.
Again, ENL at ``time point one'' is called $L_1$ and at ``time point two'' $L_2$.
Ideal surfaces and plots are flat and horizontal.

\vspace{0.9\baselineskip}\noindent
A few runs with different realizations of \bm{\Sigma} (not shown) indicate that the resulting figures and plots vary only little with \bm{\Sigma}.

\clearpage

\subsection{\hspace{1em} 3$\times$3 matrices}
\begin{figure}[h]
\centering
\includegraphics[width=0.95\columnwidth]{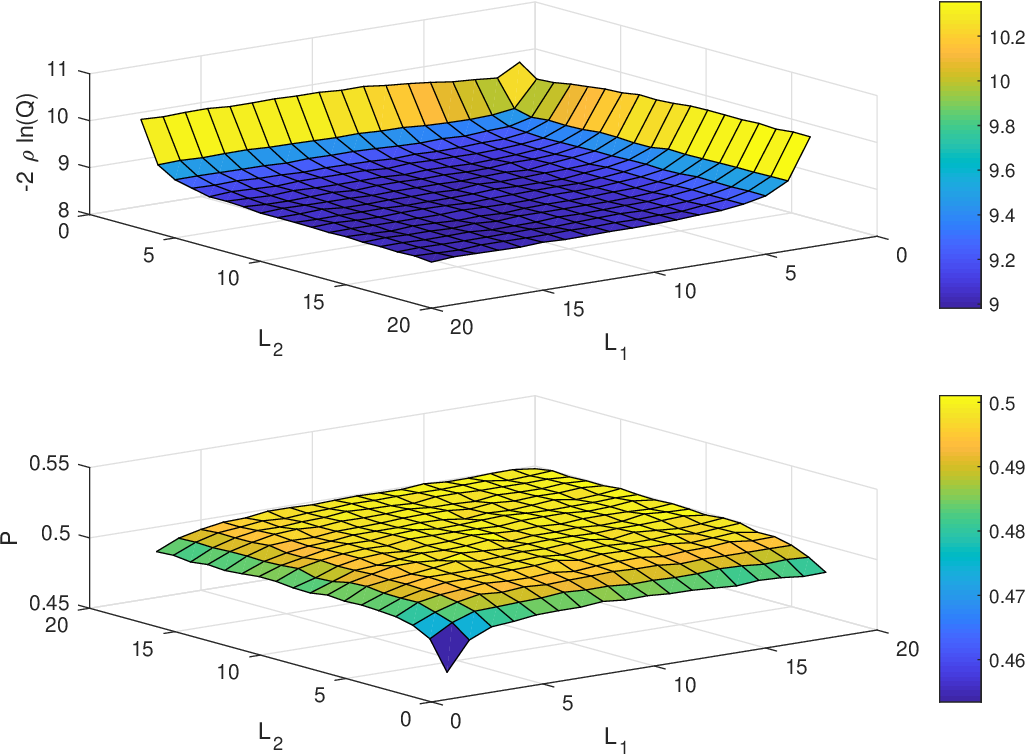}
\caption{\label{fig:QP3x3surf}Mean values of the test statistic $-2 \rho \ln Q$ (top) for
%10,000
128k
samples generated for two instances of 3$\times$3 complex Wishart distributed matrices with the same \bm{\Sigma}, i.e., a no-change situation. The test statistic is shown as a function of the number of looks at the two instances. The associated $P$-value is shown at the bottom.}
%\end{figure}
\vspace{\baselineskip}
%\begin{figure}
\centering
\includegraphics[width=0.95\columnwidth]{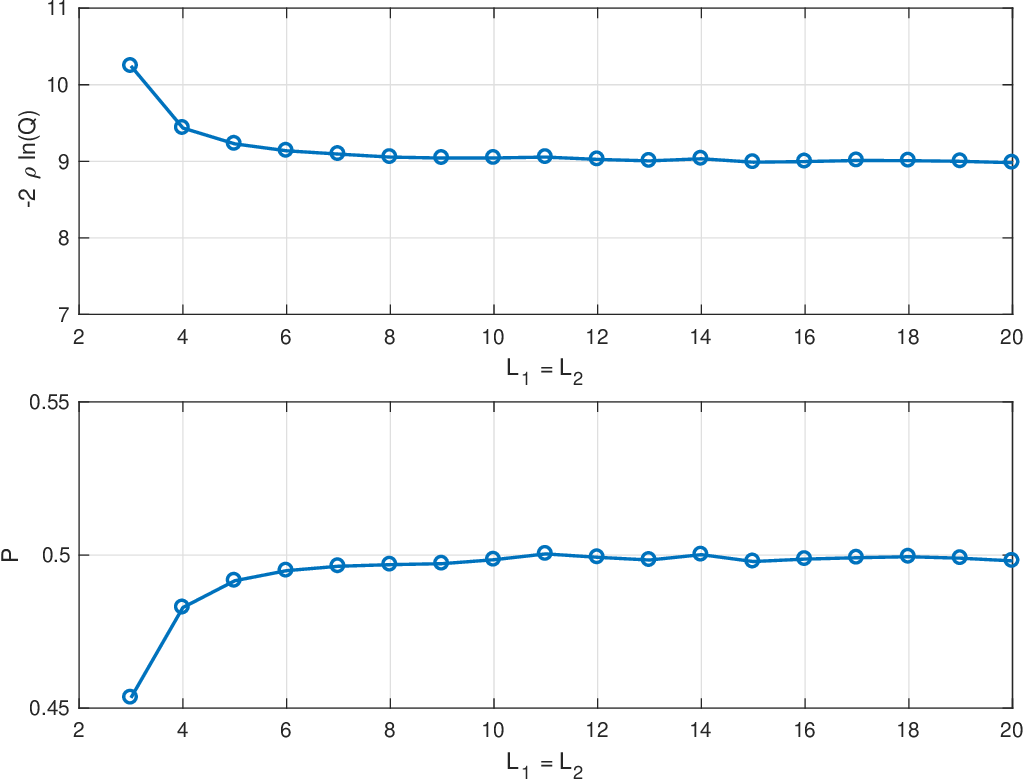}
\caption{\label{fig:QP3x3plot}Mean values of the test statistic $-2 \rho \ln Q$ (top) for
%10,000
128k
samples generated for two instances of 3$\times$3 complex Wishart distributed matrices with the same \bm{\Sigma}, i.e., a no-change situation. The test statistic is shown as a function of the same number of looks at the two instances. The associated $P$-value is shown at the bottom.}
\end{figure}

\begin{figure}[h!]
\centering
\includegraphics[width=0.95\columnwidth]{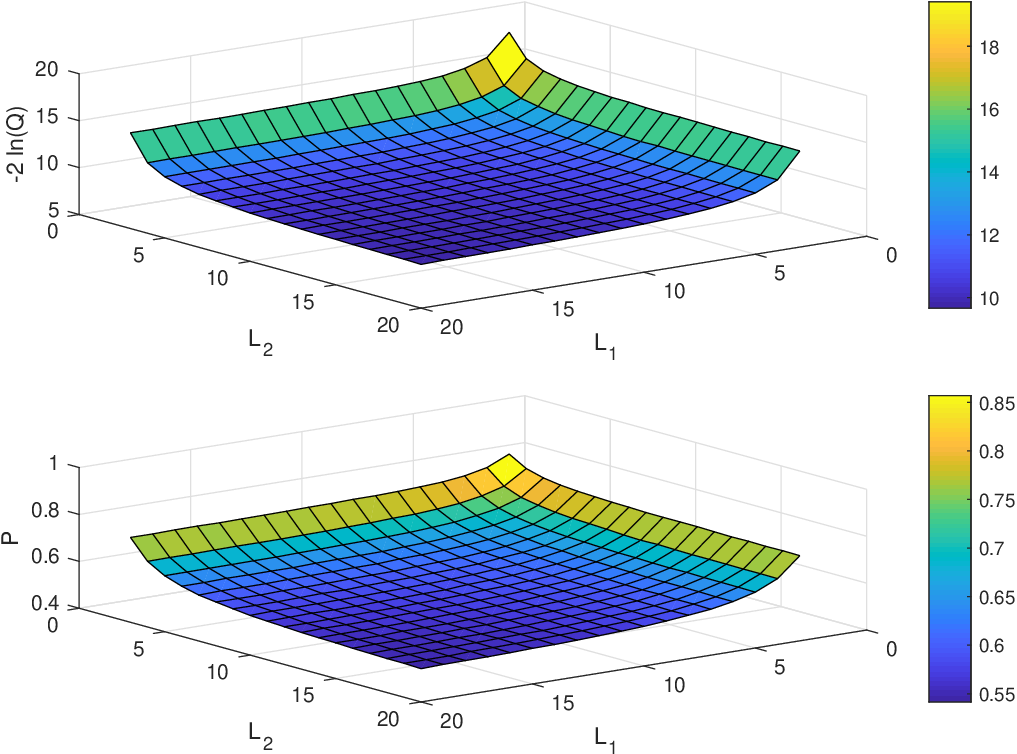}
\caption{\label{fig:QP03x3surf}Mean values of the test statistic $-2 \ln Q$ (top) for
%10,000
128k
samples generated for two instances of 3$\times$3 complex Wishart distributed matrices with the same \bm{\Sigma}, i.e., a no-change situation. The test statistic is shown as a function of the number of looks at the two instances. The associated $P$-value is shown at the bottom.}
%\end{figure}
\vspace{\baselineskip}
%\begin{figure}
\centering
\includegraphics[width=0.95\columnwidth]{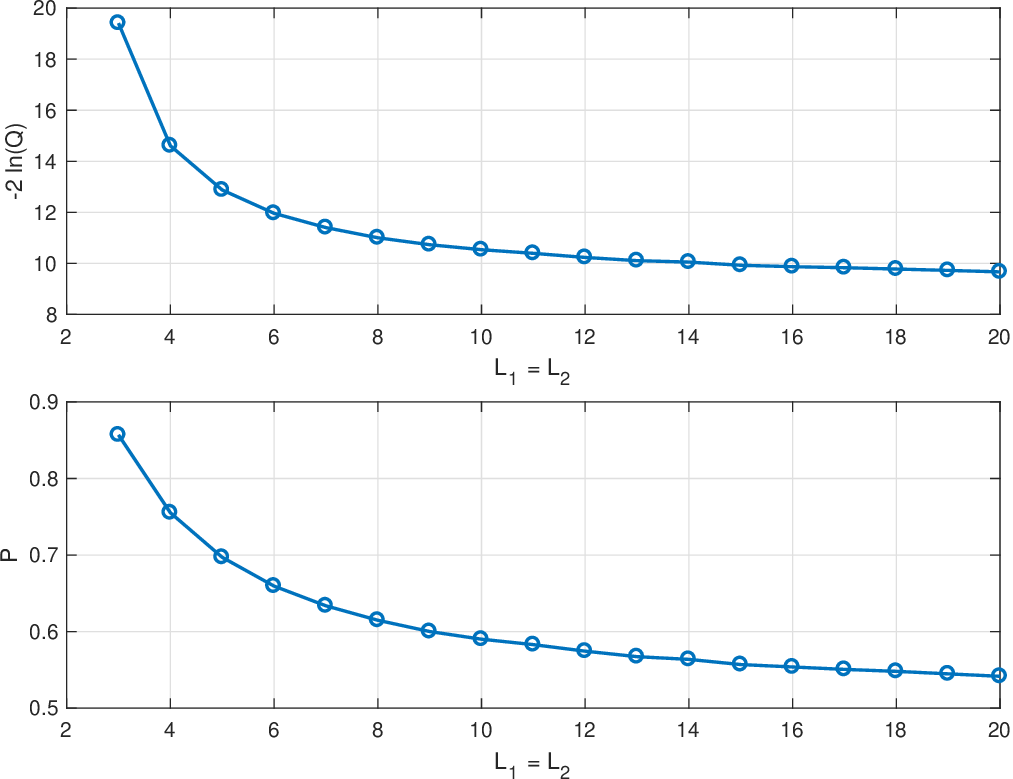}
\caption{\label{fig:QP03x3plot}Mean values of the test statistic $-2 \ln Q$ (top) for
%10,000
128k
samples generated for two instances of 3$\times$3 complex Wishart distributed matrices with the same \bm{\Sigma}, i.e., a no-change situation. The test statistic is shown as a function of the same number of looks at the two instances. The associated $P$-value is shown at the bottom.}
\end{figure}

\vspace{-0.6\baselineskip}
Figures~\ref{fig:QP3x3surf}, \ref{fig:QP3x3plot}, \ref{fig:QP03x3surf} and \ref{fig:QP03x3plot} show
mean values of the test statistics $-2 \rho \ln Q$ , $-2 \ln Q$ and associated $P$-values for
%10,000
128k
%23.5 min elapsed time
samples generated for two instances of matrices with the same \bm{\Sigma} corresponding to no-change in a change detection setting.

\vspace{0.9\baselineskip}\noindent
$-2 \rho \ln Q$ and its associated $P$-value clearly outperform $-2 \ln Q$ and its $P$-value.
For same ENL, $-2 \rho \ln Q$ and its $P$ stop varying as functions of ENL at a value of around 9.

\vspace{0.9\baselineskip}\noindent
Ideal surfaces and plots are flat and horizontal, for $-2 \rho \ln Q$ ideal values are just over 9 (depending on $\rho$ and therefore on ENL),
for $-2 \ln Q$ ideal values are 9, and
for $P$ ideal values are 1/2.

\newpage
\vspace{0.9\baselineskip}\noindent
Note the different scales and the different orientations of the $L_1$ and $L_2$ axes in
Figures~\ref{fig:QP3x3surf} to \ref{fig:QP03x3plot}.

\clearpage

\subsection{\hspace{1em} 2$\times$2 matrices}
\begin{figure}[h!]
\centering
\includegraphics[width=0.95\columnwidth]{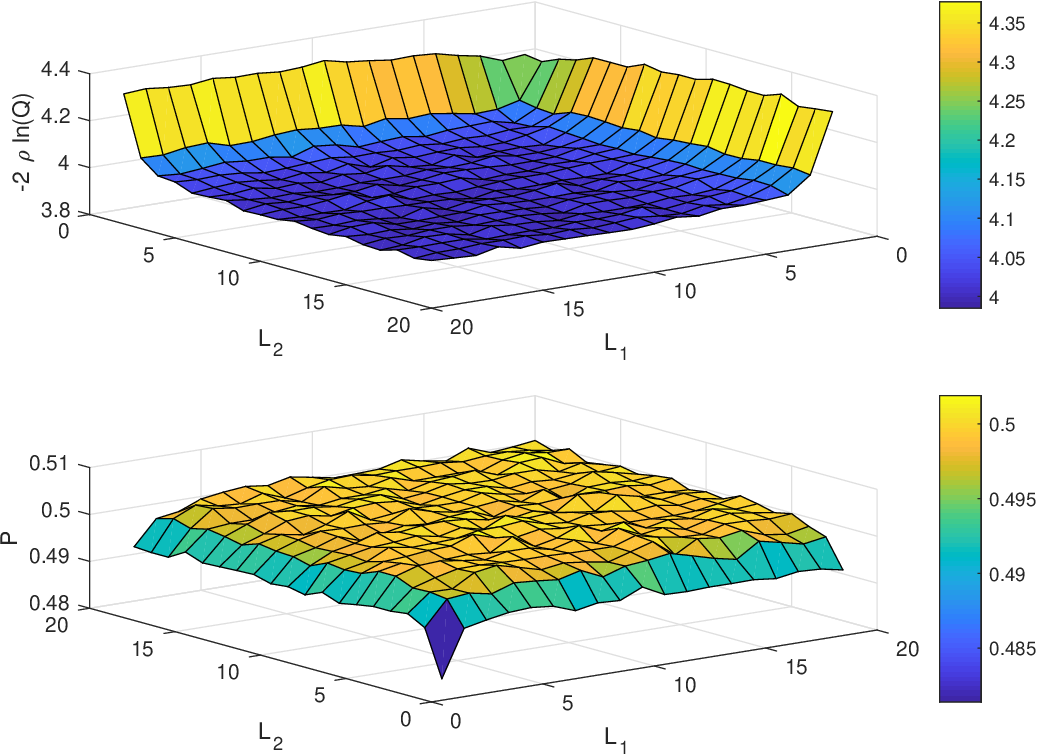}
\caption{\label{fig:QP2x2surf}Mean values of the test statistic $-2 \rho \ln Q$ (top) for
%10,000
128k
samples generated for two instances of 2$\times$2 complex Wishart distributed matrices with the same \bm{\Sigma}, i.e., a no-change situation. The test statistic is shown as a function of the number of looks at the two instances. The associated $P$-value is shown at the bottom.}
%\end{figure}
\vspace{\baselineskip}
%\begin{figure}
\centering
\includegraphics[width=0.95\columnwidth]{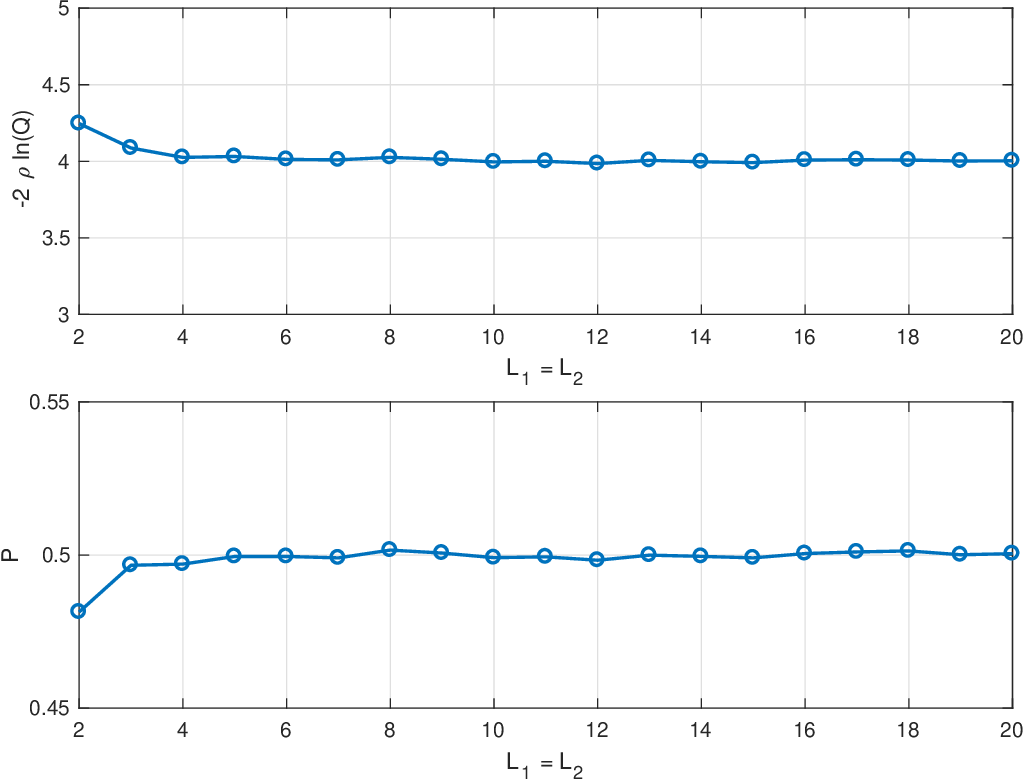}
\caption{\label{fig:QP2x2plot}Mean values of the test statistic $-2 \rho \ln Q$ (top) for
%10,000
128k
samples generated for two instances of 2$\times$2 complex Wishart distributed matrices with the same \bm{\Sigma}, i.e., a no-change situation. The test statistic is shown as a function of the same number of looks at the two instances. The associated $P$-value is shown at the bottom.}
\end{figure}

\begin{figure}[h!]
\centering
\includegraphics[width=0.95\columnwidth]{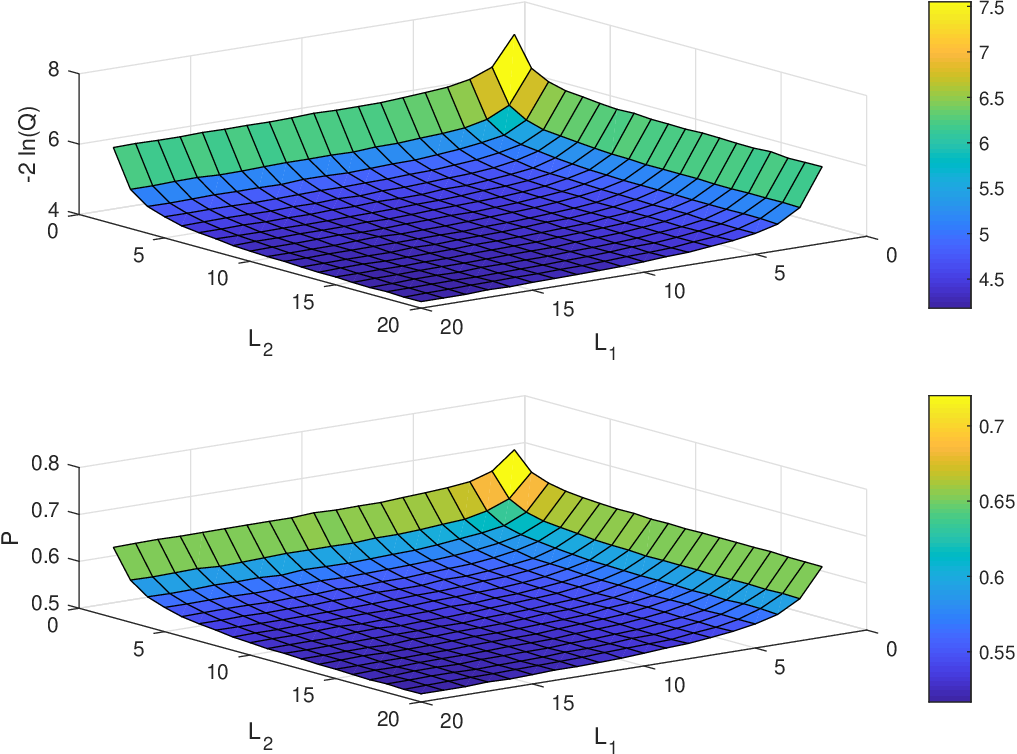}
\caption{\label{fig:QP02x2surf}Mean values of the test statistic $-2 \ln Q$ (top) for
%10,000
128k
samples generated for two instances of 2$\times$2 complex Wishart distributed matrices with the same \bm{\Sigma}, i.e., a no-change situation. The test statistic is shown as a function of the number of looks at the two instances. The associated $P$-value is shown at the bottom.}
%\end{figure}
\vspace{\baselineskip}
%\begin{figure}
\centering
\includegraphics[width=0.95\columnwidth]{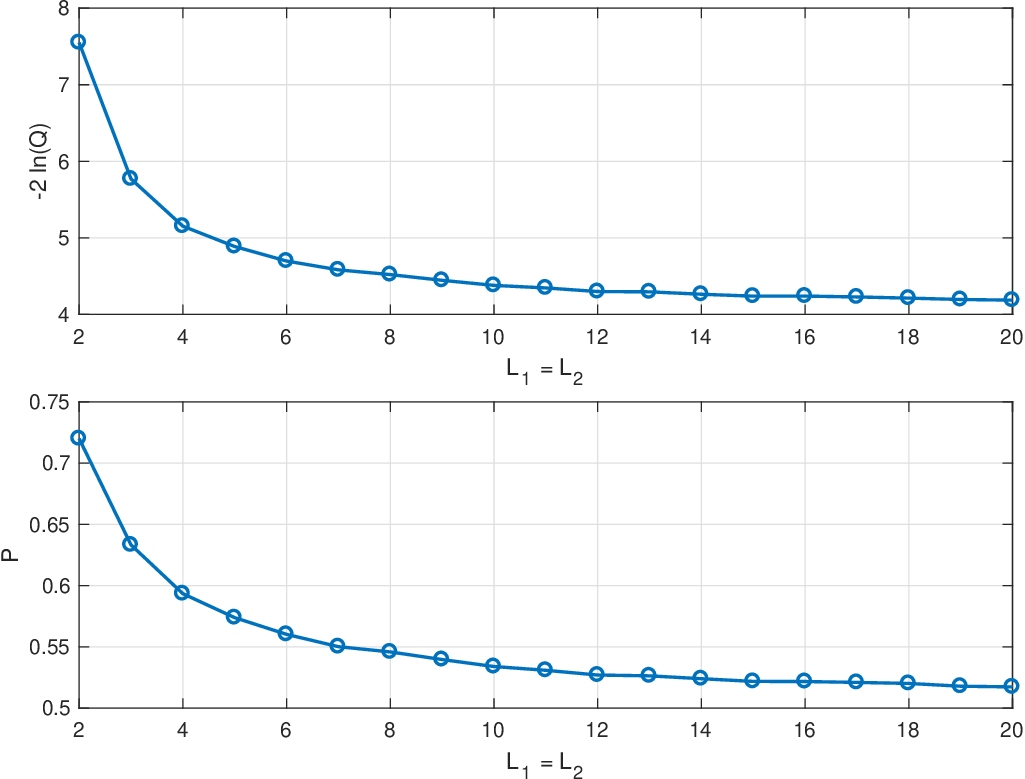}
\caption{\label{fig:QP02x2plot}Mean values of the test statistic $-2 \ln Q$ (top) for
%10,000
128k
samples generated for two instances of 2$\times$2 complex Wishart distributed matrices with the same \bm{\Sigma}, i.e., a no-change situation. The test statistic is shown as a function of the same number of looks at the two instances. The associated $P$-value is shown at the bottom.}
\end{figure}

\vspace{-0.6\baselineskip}
Figures~\ref{fig:QP2x2surf}, \ref{fig:QP2x2plot}, \ref{fig:QP02x2surf} and \ref{fig:QP02x2plot} show
mean values of the test statistics $-2 \rho \ln Q$, $-2 \ln Q$ and associated $P$-values for
%10,000
128k
samples generated for two instances of matrices with the same \bm{\Sigma} corresponding to no-change in a change detection setting.

\vspace{0.9\baselineskip}\noindent
$-2 \rho \ln Q$ and its associated $P$-value clearly outperform $-2 \ln Q$ and its $P$-value.
For same ENL, $-2 \rho \ln Q$ and its $P$ stop varying as functions of ENL at a value of around 4.

\vspace{0.9\baselineskip}\noindent
Ideal surfaces and plots are flat and horizontal, for $-2 \rho \ln Q$ ideal values are just over 4 (depending on $\rho$ and therefore on ENL),
for $-2 \ln Q$ ideal values are 4, and
for $P$ ideal values are 1/2.


\newpage
\vspace{0.9\baselineskip}\noindent
Note the different scales and the different orientations of the $L_1$ and $L_2$ axes in
Figures~\ref{fig:QP2x2surf} to \ref{fig:QP02x2plot}.

\clearpage

\subsection{\hspace{1em} 2$\times$2 matrices, diagonal only}
\begin{figure}[h!]
\centering
\includegraphics[width=0.95\columnwidth]{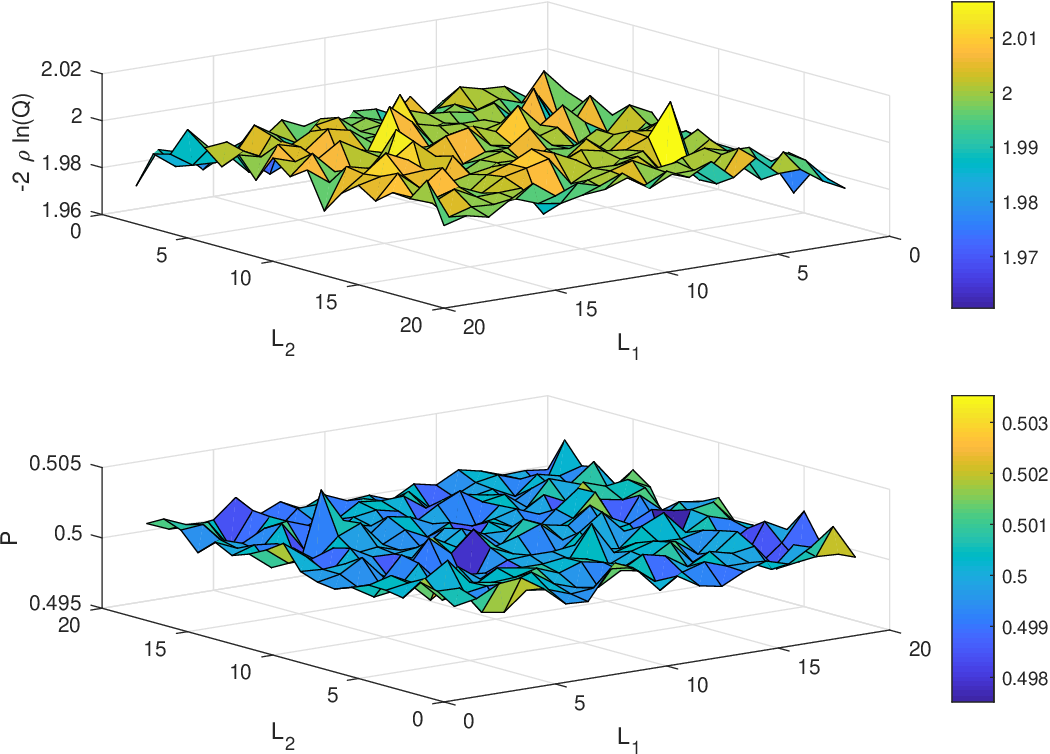}
\caption{\label{fig:QP2x2diagsurf}Mean values of the test statistic $-2 \rho \ln Q$ (top) for
%10,000
128k
samples generated for two instances of 2$\times$2 diagonal-only matrices with the same \bm{\Sigma}, i.e., a no-change situation. The test statistic is shown as a function of the number of looks at the two instances. The associated $P$-value is shown at the bottom.}
%\end{figure}
\vspace{\baselineskip}
%\begin{figure}
\centering
\includegraphics[width=0.95\columnwidth]{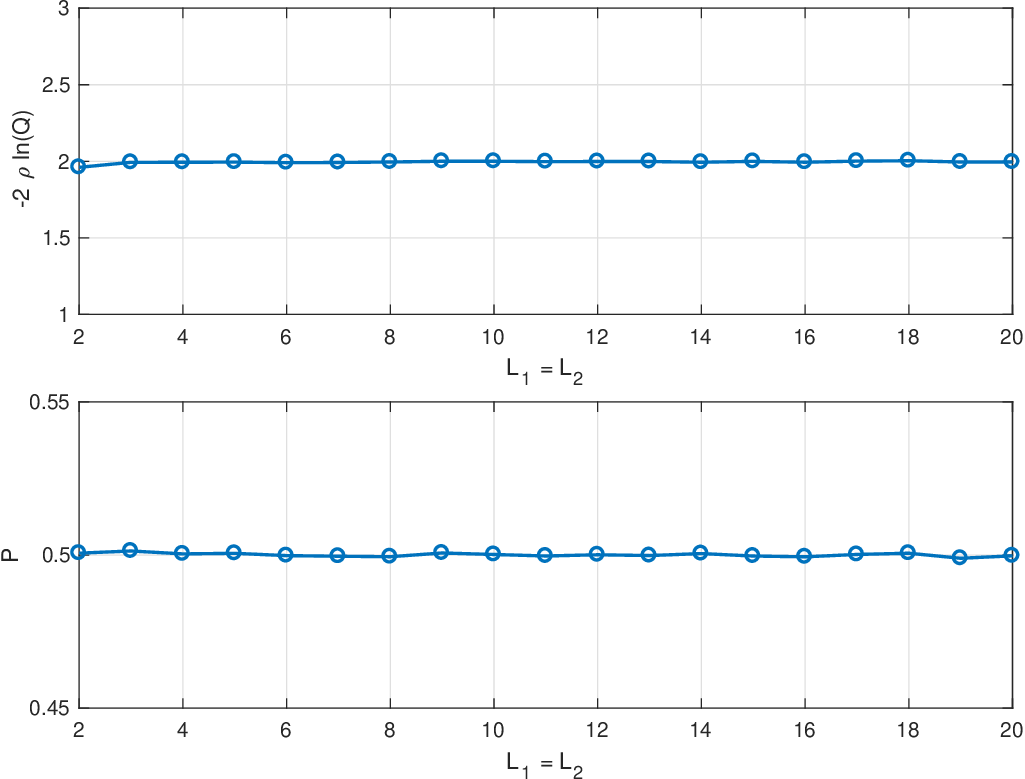}
\caption{\label{fig:QP2x2diagplot}Mean values of the test statistic $-2 \rho \ln Q$ (top) for
%10,000
128k
samples generated for two instances of 2$\times$2 diagonal-only matrices with the same \bm{\Sigma}, i.e., a no-change situation. The test statistic is shown as a function of the same number of looks at the two instances. The associated $P$-value is shown at the bottom.}
\end{figure}

\begin{figure}[h!]
\centering
\includegraphics[width=0.95\columnwidth]{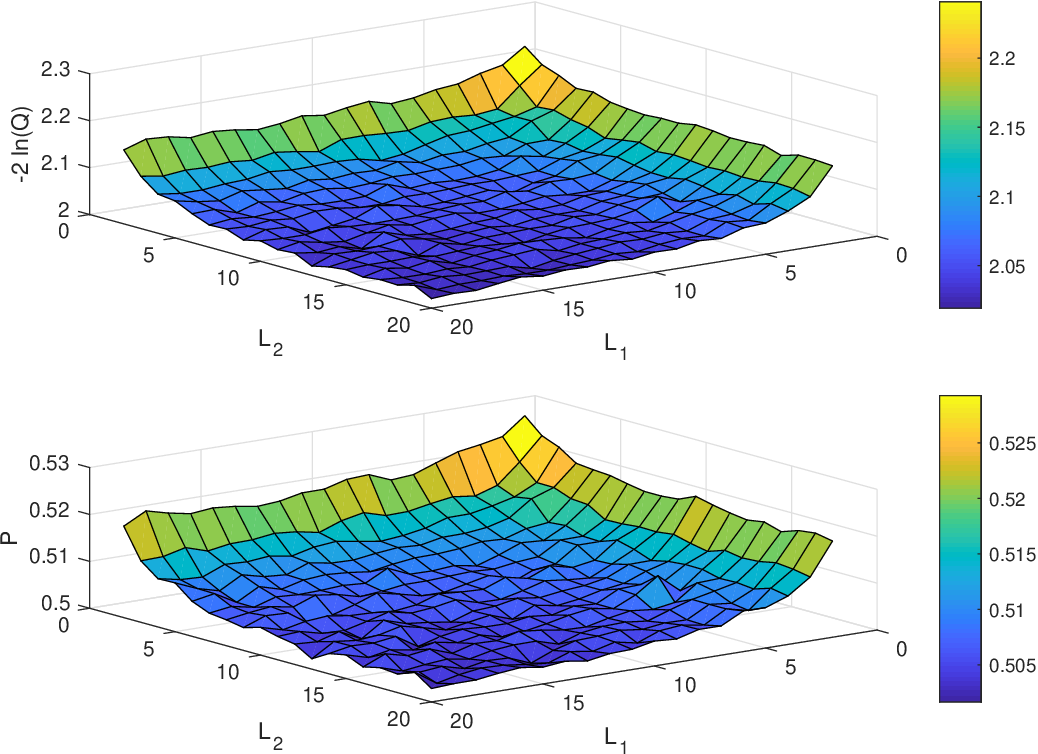}
\caption{\label{fig:QP02x2diagsurf}Mean values of the test statistic $-2 \ln Q$ (top) for
%10,000
128k
samples generated for two instances of 2$\times$2 diagonal-only matrices with the same \bm{\Sigma}, i.e., a no-change situation. The test statistic is shown as a function of the number of looks at the two instances. The associated $P$-value is shown at the bottom.}
%\end{figure}
\vspace{\baselineskip}
%\begin{figure}
\centering
\includegraphics[width=0.95\columnwidth]{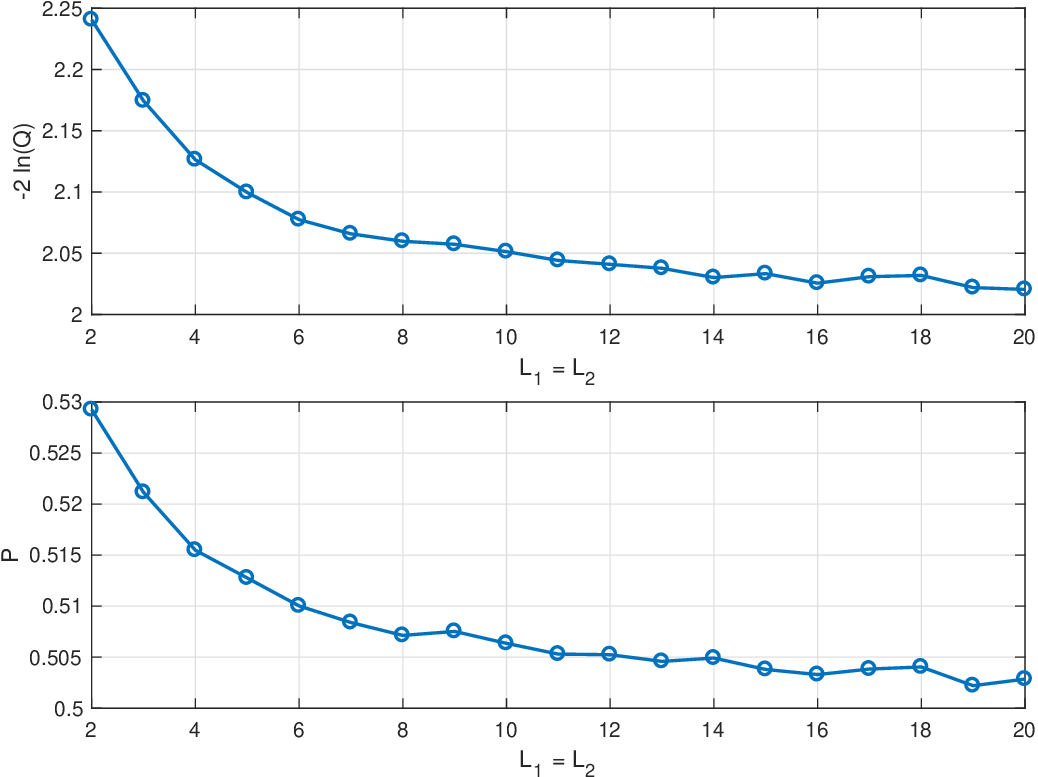}
\caption{\label{fig:QP02x2diagplot}Mean values of the test statistic $-2 \ln Q$ (top) for
%10,000
128k
samples generated for two instances of 2$\times$2 diagonal-only matrices with the same \bm{\Sigma}, i.e., a no-change situation. The test statistic is shown as a function of the same number of looks at the two instances. The associated $P$-value is shown at the bottom.}
\end{figure}

\vspace{-0.6\baselineskip}
Figures~\ref{fig:QP2x2diagsurf}, \ref{fig:QP2x2diagplot}, \ref{fig:QP02x2diagsurf} and \ref{fig:QP02x2diagplot} show
mean values of the test statistics $-2 \rho \ln Q$ , $-2 \ln Q$ and associated $P$-values for
%10,000
128k
samples generated for two instances of matrices with the same \bm{\Sigma} corresponding to no-change in a change detection setting.

\vspace{0.9\baselineskip}\noindent
$-2 \rho \ln Q$ and its associated $P$-value clearly outperform $-2 \ln Q$ and its $P$-value.
In this case, $-2 \rho \ln Q$ and its $P$ seem not to vary as functions of ENL.

\vspace{0.9\baselineskip}\noindent
Ideal surfaces and plots are flat and horizontal, for $-2 \rho \ln Q$ ideal values are just over 2 (depending on $\rho$ and therefore on ENL),
for $-2 \ln Q$ ideal values are 2, and
for $P$ ideal values are 1/2.

\newpage
\vspace{0.9\baselineskip}\noindent
Note the different scales and the different orientations of the $L_1$ and $L_2$ axes in
Figures~\ref{fig:QP2x2diagsurf} to \ref{fig:QP02x2diagplot}.
\clearpage

\section{\hspace{1em} Conclusions}
\vspace{-0.9\baselineskip}
We have used generated data to simulate no target in a target detection setting and no change in a change detection setting in (polarimetric) bi-temporal SAR data.
We have generated 3$\times$3 Hermitian, positive definite matrices, 2$\times$2 Hermitian, positive definite matrices, and
2$\times$2 diagonal-only matrices to emulate polarimetric data (from for example Radarsat-2) and dual polarimetry data (from for example Sentinel-1).

\vspace{0.9\baselineskip}\noindent
The first experiment in a target detection setting with very different number of looks at the two instances
(one instance is a (few) centre pixel(s), the other instance represents the surrounding background pixels)
gives histograms which very beautifully show the expected behaviour for no target including the uniform distribution of $P$.

%Ideal surfaces and plots of test statistics and associated $P$-values are flat and horizontal.

\vspace{0.9\baselineskip}\noindent
The second experiment shows that
$-2 \rho \ln Q$ and its associated $P$-value clearly outperform $-2 \ln Q$ and its $P$-value.
For same ENL,
%at the two time points,
$-2 \rho \ln Q$ and the associated $P$-value stop varying as functions of ENL at a value corresponding to the number of degrees of freedom.
%$-2 \rho \ln Q$ and the associated $P$-value stop varying as functions of ENL at a value of around 8-9 for 3$\times$3 matrices, around 4 for 2$\times$2 matrices, and they seem not to vary with ENL for 2$\times$2 diagonal-only matrices.

\vspace{0.9\baselineskip}\noindent
A few runs with different realizations of \bm{\Sigma} (not shown) indicate that the resulting figures and plots vary only little with \bm{\Sigma}.

\vspace{0.9\baselineskip}\noindent
Matlab code
%{\tt wishart\_change.m}
%and
{\tt wc\_rand.m}
to generate complex Wishart distributed, Hermitian, positive definite covariance matrix samples 
and to be used together with already published software to do the Wishart based change detection
\cite{wishartchange2003,wishartJSTARS2015,wishartSeveralRj2016,wishartCJRS2017},
is available on
%the first author's
Allan Nielsen's
homepage.
%\footnote{\hspace{1pt}https://people.compute.dtu.dk/alan}.

%\newpage

\raggedright

\bibliographystyle{IEEEbib}
\bibliography{../wishart}

\end{document}